\documentclass[10pt]{article}
\usepackage{amsmath}
\usepackage{amsfonts}
\usepackage{amssymb}
\usepackage{color}
\usepackage[english]{babel}
\usepackage{xcolor}
\usepackage{graphics}
\usepackage{graphicx}
\begin{document}
\title{
On the velocity of moving relativistic unstable quantum systems
}
\author{K. Urbanowski\footnote{e--mail: K.Urbanowski@if.uz.zgora.pl}\\
Institute of Physics,
University of Zielona G\'{o}ra,  \\
ul. Prof. Z. Szafrana 4a,
65-516 Zielona G\'{o}ra,
Poland}
\maketitle

\begin{abstract}
We study properties  of moving relativistic quantum
unstable systems. We show that in contrast to the properties of classical
particles and quantum stable objects
the velocity of moving freely relativistic quantum unstable systems
can not be constant in time. We show that this new quantum effect  results
from the fundamental principles of the quantum theory and physics: It is a
consequence of the principle of conservation of energy and of the fact that the
mass of the quantum unstable system is not defined. This effect
can affect the form of the decay law of moving relativistic quantum unstable systems.
\end{abstract}

\maketitle

\section{Introduction}

Physicists studying the decay processes of  unstable quantum systems
moving with the velocity $\vec{v}$ relative to the rest reference frame of an observer,
and trying to derive theoretically the decay law of such systems are confronted
with the following problem: Which of the two possible assumptions:
$\vec{v}= const$ , or perhaps
$\vec{p} = const$ (where $\vec{p}$ is the momentum of the
moving unstable system), will get decay law correctly
describing the real properties of such system.
When one considers classical physics decay processes
the mentioned assumptions both lead to the decay law of the same form. Namely
from the standard, text book  considerations one finds that if
the decay law of the unstable particle in rest
 has the exponential form ${\cal P}_{0}(t) = \exp\,[- {\it\Gamma}_{0}\,t]$
then the decay law of the moving particle with momentum $\vec{p} \neq0$ is
${\cal P}_{p}(t)\,={\cal P}_{0}(t/\gamma)\,\equiv\,
\exp\,[-\,\frac{{\it\Gamma}_{0}\,t}{\gamma}] \equiv {\cal P}_{v}(t)$, where $t$ denotes time,
${\it\Gamma}_{0}$ is the decay rate (time $t$ and ${\it\Gamma}_{0}$
are measured in the rest reference frame of the particle),
$\gamma$ is the
relativistic Lorentz factor, $\gamma \equiv 1 / {\sqrt{1 - v^{2}}}$,
$v=|\vec{v}|$ and ${\cal P}_{v}(t)$
is the decay law of the particle moving with the constant velocity $\vec{v}$.
(We use $\hbar = c = 1$ units, and thus  $v< 1$).
It is almost common belief that this equality is valid also for any $t$
in the case of quantum decay processes
and does not depend on the model of the unstable system considered.
The cases $\vec{p} =const$ and $\vec{v}=const$ both were studied in the literature.
The assumption $\vec{p}= const$ was used in \cite{stefanovich,shirkov} to
derive the survival probability ${\cal P}_{p}(t)$. From these studies it follows that
in the case of moving quantum unstable systems the
relation ${\cal P}_{p}(t) \simeq {\cal P}_{0}(t/\gamma)$ is
valid to a sufficient accuracy only for not more than a few lifetimes and
that for times much longer than a few lifetimes ${\cal P}_{p}(t) > {\cal P}_{0}(t/\gamma)$
(see \cite{shirkov,ku-2014}).
The assumption $\vec{v} = const$ was used, eg. in \cite{exner} to derive
the decay law of moving quantum unstable systems. Unfortunately the result
obtained in \cite{exner} is similar to the case $\vec{p} = const$: ${\cal P}_{v}(t) \simeq
{\cal P}_{0}(t/\gamma)$ only for no more than a few lifetimes.
What is more it appears
that the assumption  $\vec{v}=const$ may lead to the relation ${\cal P}_{v}(t) = {\cal P}_{0}(\gamma\,t)$,
i.e. to the result never observed in experiments \cite{shirkov1}.

Unfortunately
the experiments did not give any decisive answer for the problem
which is the correct assumption: $\vec{p} = const$ or $\vec{v} =const$?
It is because
all known tests of the relation ${\cal P}_{v(p)}(t) \simeq {\cal P}_{0}(t/\gamma)$
 were performed
for times $t \sim \tau_{0}$ (where $\tau_{0}$ is the lifetime)
(see, eg. \cite{muon1,muon2}). Note that the same relation
obtained in \cite{stefanovich,shirkov,exner} is
approximately valid for the same times $t$ (see also discussion in \cite{giacosa1}).
The problem seems to be extremely important in accelerator physics where the
correct interpretation of the obtained results depends on  knowledge of the properly calculated
decay law of the moving unstable particles created in the collisions observed.
Similarly the proper interpretation of  results of observations of astrophysical
processes in which a huge numbers of elementary particles
(including unstable one) are produced is impossible without knowing the correct
form of the decay law of unstable particles created in these processes.
So the further theoretical studies of the above described problem
are necessary and seems to be important.

In this letter we analyze general properties of unstable quantum system
from the point of view of fundamental principles of physics and quantum theory.
Here we show that the principle of the conservation of the energy does not allow
any moving quantum unstable system to move with the velocity $\vec{v}$ constant in time.

\section{ Quantum unstable  systems}

\begin{figure}[h!]
\begin{center}
\includegraphics[width=70mm]{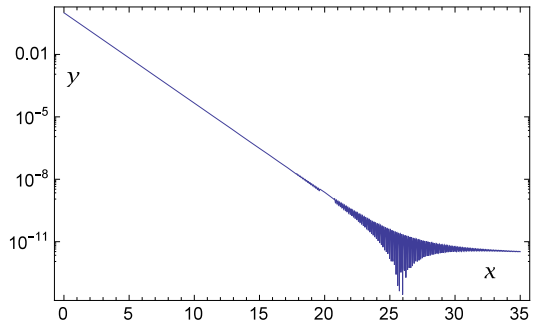}\\
\caption{Decay curve obtained for $\omega(m)$ given by Eq. (\ref{omega-BW}).
Axes: $y = {\cal P}_{0}(t)$, the logarithmic scale; $x =t / \tau_{0} $. (Time is measured in lifetimes).}
 \label{f1}
\end{center}
\end{figure}

The main information about properties of  quantum unstable   systems
is contained in their decay law, that is in their survival probability.
Let the reference frame ${\cal O}$ be the common inertial rest
frame for the observer and for the unstable system.
Then  if one knows that the system in the rest frame is in the initial unstable
state $|\phi\rangle$, which was prepared at the initial instant $t_{0} =0$,
one can calculate
its survival probability, ${\cal P}_{0}(t)$, which equals
${\cal P}_{0}(t) = |a(t)|^{2}$,  where
$a(t)$ is the survival amplitude, $a(t) = \langle \phi|\phi;t\rangle$,
and $|\phi;t\rangle = e^{\textstyle{-itH}}\,|\phi\rangle$, $H$ is the
total selfadjoint Hamiltonian of the system under considerations,
$|\phi \rangle, |\phi;t\rangle
\in {\cal H}$ and ${\cal H}$ is the Hilbert space of states of
the considered system.
So in order to calculate the amplitude $a(t)$ one should know the state $|\phi\rangle$.
Within the standard approach the unstable state $|\phi\rangle$ is modeled
 as the  following wave--packet \cite{fock,khalfin,fonda}
\begin{eqnarray}
 |\phi\rangle
= \int_{\mu_{0}}^{\infty}c(m)\, |m\rangle\,dm,\label{phi-1}
\end{eqnarray}
where  $\mu_{0}$ is the lower bound of the the continuous part  $\sigma_{c}(H)$ of the spectrum of  $H$,
and vectors $|m\rangle $ solve the equation
\begin{equation}
H|m\rangle = m\,|m\rangle,
\;\;\;m\in \sigma_{c}(H). \label{H-m}
\end{equation}
Eigenvectors $|m\rangle$ are normalized as follows
\begin{equation}
\langle m|m'\rangle = \delta(m - m').
\end{equation}
We  require the state $|\phi\rangle$  to be normalized:
So it has to be $\int_{\mu_{0}}^{\infty}|c(m)|^{2}\,dm = 1$.
Thus
\begin{equation}
|\phi; t\rangle = e^{\textstyle{-itH}}\,|\phi\rangle
\equiv \int_{\mu_{0}}^{\infty}c(m)\,e^{\textstyle{-itm}} |m\rangle\,dm,\label{phi-t-1}
\end{equation}
which allows one
to represent the amplitude
$a(t)$ as the Fourier transform
of the mass (energy) distribution function,
$\omega(m) \equiv |c(m)|^{2}$,
\begin{equation}
a(t) = \int\,\omega(m)\,e^{\textstyle{-itm}}\,dm, \label{a(t)-omega}
\end{equation}
where $\omega (m) \geq 0$ and $\omega (m) = 0$ for $m < \mu_{0}$
\cite{fock,khalfin,fonda,muga,muga-1,nowakowski,krauss,giraldi}
(see also \cite{stefanovich,shirkov,exner,ku-2014,shirkov1}).
From the last relation and from the Rieman--Lebesque lemma it follows that
$|a(t)| \to 0$ as $t \to \infty$. It is because from the normalization
condition $\int_{\mu_{0}}^{\infty}|c(m)|^{2}\,dm \equiv \int_{\mu_{0}}^{\infty} \,
\omega (m)\,dm = 1$ it follows that $\omega (m)$ is an absolutely integrable function.
The typical form of the survival probability ${\cal P}_{0}(t)$ is presented in Fig (\ref{f1}),
where the calculations were performed for $\omega (m)$ having the Breit--Wigner form,
\begin{equation}
\omega_{BW} ({m})
 \equiv \frac{N}{2\pi}\,  {\it\Theta} ( m - \mu_{0}) \
\frac{{\it\Gamma}_{\phi}^{0}}{({ m }-{ M}_{\phi}^{0})^{2} +
({{\it\Gamma}_{\phi}^{0}} / {2})^{2}},    \label{omega-BW}
\end{equation}
assuming for simplicity that $s_{0} = ({M_{\phi}^{0}} - \mu_{0})/ {{\it\Gamma}_{\phi}^{0}} = 50$.
Here ${\it\Theta}(m)$ is a step function: ${\it\Theta}(m) = 0\;\;{\rm  for}\;\; m \leq 0$
and ${\it\Theta}(m) = 1\;\;{\rm for}\;\; m>0  $.

Note that
\begin{equation}
H|\phi\rangle \equiv \int_{\mu_{0}}^{\infty}\,m\,c(m)\, |m\rangle\,dm,
\label{H-phi}
\end{equation}
which means that the vector $|\phi\rangle$ corresponding to an unstable state
is not the eigenvector for the Hamiltonian $H$. In other words
in the rest frame considered there does not exist any
number, let us denote it by $m_{\phi}^{0}$, such that it would
be $H |\phi\rangle = m_{\phi}^{0}|\phi \rangle$.
This means that the mass (that is the rest mass $m_{\phi}^{0}$) of the unstable
quantum system described by the vector $|\phi\rangle$ is not defined.
What is more in such a case the mass of this system can not be constant in time in the state considered.
 Simply the mass  of the unstable system  can
 not take the exact constant value in the state $|\phi\rangle$
otherwise it would not be any decay, that is it
would be ${\cal P}_{0}(t) \equiv |\langle \phi |\exp\,[-itH]|\phi \rangle|^{2} = 1$ for all $t$.
In general such quantum systems are characterized by
the time independent mass (energy) distribution density $\omega (m)$,
that is by the modulus of the expansion coefficient $c(m)$, but
not by the exact value of the mass.
In  this case instead of the mass,  the average mass,
$\langle m_{\phi}\rangle$, of the unstable system
can be determined knowing $\omega (m)$
or the instantaneous mass of this system \cite{giraldi,ku-2009,ku1-2009,ku1-2014}.
The average mass is defined by means  of the standard formula:
$\langle m_{\phi}\rangle = \int_{\mu_{0}}^{\infty}\,m\,\omega(m)\,dm$.
The instantaneous mass $m_{\phi}^{0}(t)$ (energy) can be found used using
the exact effective Hamiltonian $h_{\phi}(t)$ governing the time evolution
in the subspace of states spanned by the vector $|\phi\rangle$,
\begin{eqnarray}
h_{\phi}(t) &=& \frac{i}{a(t)}\,\frac{\partial a(t)}{\partial t}, \label{h(t)-1}\\
&\equiv&  \frac{\langle \phi |H|\phi;t\rangle}{\langle \phi |\phi;t\rangle},
\label{h-equiv}
\end{eqnarray}
which results from the Schr\"{o}dinger equation when one looks for the exact
evolution equation for the mentioned subspace of states
(for details see \cite{pra,giraldi,ku-2009,ku1-2009,ku1-2014}).
Within the assumed system of units the instantaneous mass (energy) of the unstable quantum system in
the rest reference frame is  the real part of $h_{\phi}(t)$:
\begin{equation}
m_{\phi}^{0}(t) = \Re\,[h_{\phi}(t)], \label{m(t)}
\end{equation}
and ${\it\Gamma}_{\phi} (t) = - 2\Im\,[h_{\phi}(t)]$ is the instantaneous decay rate.

\begin{figure}[h!]
\begin{center}
\includegraphics[width=70mm]{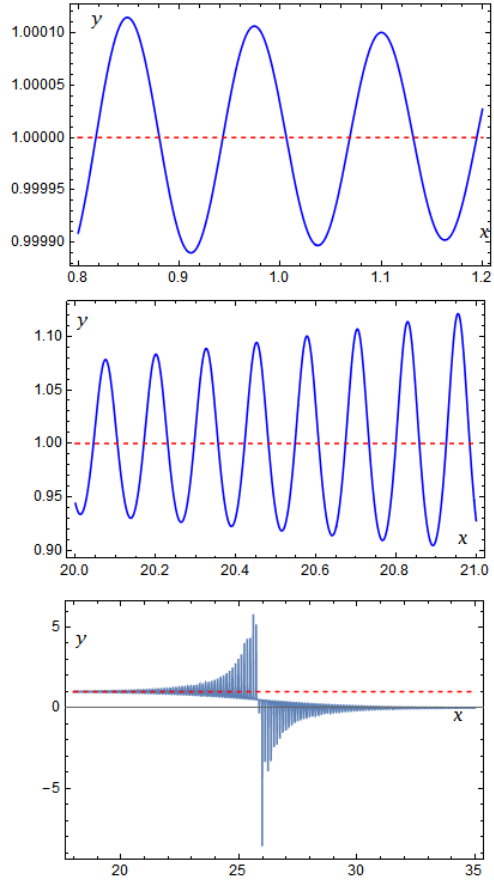}\\
\caption{A typical form of the instantaneous mass $m_{\phi}(t)$ as
a function of time obtained for $\omega_{BW}(m)$.
Axes: $y = \kappa (t)$, where $\kappa(t)$ is defined  by (\ref{kappa});
$x =t / \tau_{\phi} $: Time is measured in lifetimes.
The horizontal dashed line represents the value of $m_{\phi}^{0}(t) = M_{\phi}^{0}$}
  \label{f2}
\end{center}
\end{figure}

Using the relation (\ref{h-equiv}) one can find some general properties of $h_{\phi}(t)$ and $m_{\phi}^{0}(t)$.
Indeed, if to rewrite the numerator
of the righthand side of (\ref{h-equiv}) as follows,
\begin{equation}
\langle \phi|H|\phi;t\rangle \equiv \langle
\phi|H|\phi\rangle\,a(t)\,+\,\langle \phi |H|\phi;t \rangle_{\perp}, \label{perp}
\end{equation}
where $|\phi; t\rangle_{\perp} = Q|\phi;t\rangle$,
$Q = \mathbb{I} - P$ is the projector onto the subspace od decay products,
$P = |\phi\rangle\langle \phi|$ and $\langle \phi|\phi;t\rangle_{\perp} = 0$,
then one can see
that there is a permanent contribution of
decay products described by $|\phi;t \rangle_{\perp}$ to the
instantaneous mass (energy) of the unstable state considered.  The intensity of this contribution depends on time $t$.
Using (\ref{h-equiv}) and (\ref{perp}) one finds that
\begin{equation}
h_{\phi}(t) = \langle \phi|H|\phi\rangle\, +\, \frac{\langle \phi |H|\phi; t\rangle_{\perp}}{a(t)}. \label{h-perp-1}
\end{equation}
From this relation one can see that $h_{\phi}(0) = \langle \phi|H|\phi\rangle$ if the
matrix elements $\langle \phi|H|\phi\rangle$ exists. It is because
$|\phi (t=0)\rangle_{\perp} =0$ and $a(t=0)=1$. Now let us assume that $\langle \phi|H|\phi\rangle$ exists and
$i \frac{\partial a(t)}{\partial t} \equiv \langle \phi| H|\phi;t\rangle$ is a continuous function of time $t$ for $0 \leq t < \infty$.
If these assumptions are satisfied then $h_{\phi}(t)$ is a continuous function of time $t$ for $0 \leq t < \infty$ and $h_{\phi}(0) =
\langle \phi|H|\phi\rangle$ exists.
Now if to  assume
that for $0\leq t_{1} \neq t_{2}$ there is $\Re\,[h_{\phi}(0)] = \Re\,[h_{\phi}(t_{1})] = \Re\,[h_{\phi}(t_{2})] = const$
then from the continuity of $h_{\phi}(t)$ immediately follows that
there should be $\Re\,[h_{\phi}(t)] = h_{\phi}(0) \equiv \langle \phi|H|\phi\rangle = const$ for any $t \geq 0$. Unfortunately such an observation contradicts implications of  (\ref{h-perp-1}): From this relation it follows that $\Re\,[\frac{\langle \phi |H|\phi;t\rangle_{\perp}}{a(t)}] \neq 0$ for $t > 0$ and thus $\Re\,[h_{\phi}(t > 0)] \neq \langle \phi|H|\phi\rangle \equiv \Re\,[h_{\phi}(0)]$ which shows that $m_{\phi}^{0}(t) \equiv \Re\,[h_{\phi}(t)]$
can not be constant in time. Results of numerical calculations presented in Fig (\ref{f2}) (or those one can find in \cite{ku1-2014}) confirm this conclusion.

In the general case the  mass (energy) distribution function
$\omega(E)$
has properties similar to the scattering amplitude, i.e., it can be
decomposed into a threshold factor, a pole-function $P(m)$ with a simple
pole at $m = M_{\phi}^{0} - \frac{i}{2}\,{\it\Gamma}_{\phi}^{0}$ (often modeled by a Breit-Wigner)
and a smooth form factor $F(m)$. So there is (see, eg. \cite{fonda}),
\begin{equation}
\omega(m)= {\it\Theta}(m-\mu_{0})\,(m-\mu_{0})^{\lambda + l}\,P(m)F(m), \label{omeha(m)}
\end{equation}
where $l$ is the angular momentum, $1 > \lambda \geq 0$.   In such a case
\begin{equation}
h_{\phi}^{0}(t) \simeq M_{\phi}^{0} - \frac{i}{2}\,{\it\Gamma}_{\phi}^{0},
\;\;\;( t \sim\tau_{\phi}), \label{h(t)-canonical}
\end{equation}
at canonical decay times,
that is when the survival probability has the exponential
form (here $\tau_{\phi}$ is the lifetime),
and
\begin{equation}
m_{\phi}^{0}(t) = \Re\,[h_{\phi}(t)] \simeq M_{\phi}^{0} =
\langle m_{\phi}\rangle + \Delta m^{0}_{\phi}, ( t \sim\tau_{\phi}), \label{m-canonical}
\end{equation}
at these times to a good accuracy (see \cite{pra,ku-2009,ku1-2009}). The
parameters $M_{\phi}^{0}$ and ${\it\Gamma}_{\phi}^{0}$ are the quantities
that are measured in decay and scattering experiments.
If the state vector $|\phi\rangle = |\phi_{\alpha}\rangle$ is an eigenvector
for $H$ corresponding to the eigenvalue $m_{\alpha}$ then there is
$h_{\phi_{\alpha}}(t) \equiv m_{\alpha}$.
Beyond the canonical decay times $m_{\phi}^{0}(t)$ differs from $M_{\phi}^{0}$
significantly (for details see \cite{ku-2009,ku1-2009,ku1-2014}). At
canonical decay times, values of $m_{\phi}^{0}(t)$
fluctuate (faster or slower) around $M_{\phi}^{0}$. One can
see a typical behavior  of $m_{\phi}^{0}(t)$  in
Fig (\ref{f2}), where the function,
\begin{equation}
\kappa (t) = \frac{m_{\phi}^{0}(t) - \mu_{0}}{M_{\phi}^{0} - \mu_{0}}, \label{kappa}
\end{equation}
is presented.
These results
were obtained numerically for the Breit--Wigner mass (energy) distribution function
$\omega(m) = \omega_{BW}(m)$ and for $s_{0} = 50$.
From Fig. (\ref{f2}) one can see that fluctuations of $m_{\phi}^{0}(t)$ take place at all stages
of the time evolution of the quantum unstable system. At times of order
of the lifetime, $t \sim \tau_{\phi}$, and shorter
their amplitude is so small that their impact on results of the
mass (energy) measurements can be neglected (see (\ref{m-canonical})).
With increasing time their amplitude grows up to the maximal values,
which take place at the transition times, that is when
the late time nonexponential deviations of the survival probability,
${\cal P}_{0}(t)$, begin to dominate. Thus with the increasing time, for $t > \tau_{\phi}$,
the impact of these fluctuations on behavior of the quantum unstable systems increases.

Now let us consider the case when the unstable quantum system is moving
with a velocity $\vec{v}$ relative to reference frame $\cal O$.
It is obvious that an unstable quantum system moving with the relativistic
velocity does not  turn into a classical system
but still subjects to the laws of quantum physics. So when one searches for
properties of such systems the implications following from
rules of the quantum theory are decisive.
Let us assume
that this quantum object is moving freely with the constant velocity $\vec{v}$,
\begin{equation}
\vec{v} = const,\label{v=const}
\end{equation}
 and let us admit that
$\vec{v}$ is so large that the relativistic effects can take place.
The energy $E_{\phi}$ of the quantum unstable system described in the rest
frame  by vector $|\phi\rangle$ and moving with the constant velocity $\vec{v}$
can be expressed within the system of units used as follows
\begin{equation}
E_{\phi} = m_{\phi}^{0} \,\gamma,
\label{E}
\end{equation}
where $m_{\phi}^{0}$ is the mass parameter (i.e., the rest mass) and $\gamma = const$.
Thus knowing the energy
$E_{\phi}$ and the velocity $\vec{v}$, that is the Lorentz
factor $\gamma$, one can determine the mass parameter $m_{\phi}^{0}$.

From the fundamental principles it follows that the total energy  of the moving
freely objects both quantum and classical, stable and unstable, must
be conserved.
This means that if  an experiment indicates the energy,
$E$, of such an object to be equal $E(t_{1})= E_{\phi} $ at an instant $t_{1}$ then
at any instant $t_{2} > t_{1}$ there must be $E(t_{2})= E(t_{1}) \equiv E_{\phi}$.
Now if the energy, $E$, of the moving quantum unstable system is conserved, $E = E_{\phi} =const$,
then from the assumption  it trivially results that there must be:
\begin{equation}
m_{\phi}^{0} = \frac{E_{\phi}}{\gamma} \,\equiv\, const. \label{m=const}
\end{equation}
This observation concerns also the instantaneous mass $m_{\phi}^{0}(t)$: If it was $E_{\phi} = E_{\phi}(t)$,
it would be $E_{\phi}(t)/\gamma = m_{\phi}^{0}(t)$.
The conservation of the energy
means that at any instant of time the energy has the same value,
so there must be $E_{\phi}(t) \equiv E_{\phi} = const$.
Therefore if the energy is conserved and assumption (\ref{v=const}) holds then
there must be $m_{\phi}^{0}(t) = E_{\phi}(t)/\gamma \equiv E_{\phi}/ \gamma = const$.

On the basis of this analysis one can conclude that
the rest mass $m_{\phi}^{0}$ as well as the instantaneous mass $m_{\phi}^{0}(t)$
of the moving quantum unstable system are
constant at all instants of time $t$.
But, unfortunately,
such a conclusion is in sharp contrast to the conclusion
following from the relation (\ref{H-phi}) and its consequences.
This means that one should consider the following possible situations:
Either (a) conclusions following from the quantum theoretical treatment of the
problem are  wrong (i. e. the quantum theory is wrong), or (b)  the energy
of moving quantum unstable systems is not conserved (that is, the principle
of the conservation of the energy does not apply to moving quantum unstable
systems), or simply (c) the assumption (\ref{v=const}) can not be realized
in the case of moving quantum unstable systems.
The probability that the situations (a) or (b) occur is rather negligible
small. So the only reasonable conclusion is that the case (c) take place.

This situation has a simple explanation.
Namely despite the conclusions resulting from the relation (\ref{H-phi})
in  experiments with unstable particles one observes them as
massive objects. This is not in contradiction with
the implications of the relation (\ref{H-phi}).
The conclusion that the mass of the unstable quantum system (that is the rest
mass of such a system) can not be defined and constant in time means
that in the case of such system  only the instantaneous mass varying in time
can be considered: It can only be $m_{\phi}^{0} \equiv m_{\phi}^{0}(t)$.

So in the case of the moving quantum unstable system the principle
of the conservation of the energy takes the following form:
\begin{equation}
E_{\phi} = m_{\phi}^{0}(t)\,\gamma (\vec{v}(t)) = const. \label{E=mg(v)}
\end{equation}
Thus the energy conservation forces to compensate for changes in the
instantaneous mass $m_{\phi}^{0}(t)$ by appropriate changes of the
velocity $\vec{v} \equiv \vec{v}(t)$  so that the product $m_{\phi}^{0}(t)\,
\gamma (\vec{v}(t))$ was fixed and constant in time: For any times
$t_{1} > t_{2}$ (in general $t_{1} \neq t_{2}$) there must
be $m_{\phi}^{0}(t_{1})\,\gamma (\vec{v}(t_{1})) =
m_{\phi}^{0}(t_{2})\,\gamma (\vec{v}(t_{2})) = E_{\phi}$.
(It is a pirouette like effect).
In other words
the principle of the conservation of the energy does not allow any
moving quantum unstable system to move with the velocity $\vec{v}$ constant in time.

\section{Concluding remarks}

The above conclusions result from the basic principles of the quantum theory.
Taking this into account one should  consider a possibility that in
the case of moving quantum unstable systems the assumption (\ref{v=const})
may lead to the wrong conclusions. In general the relation ${\cal P}_{v}(t) =
{\cal P}_{0}(t/\gamma)$ can be considered only as the approximate one and it can
not pretend to be rigorous. So instead of using such a relation one should rather
look for an effective formulae for the survival probability of the moving quantum
unstable systems. Such the effective formula could be obtained, e.g.,  by
replacing $\gamma = const$ in ${\cal P}_{0}(t/\gamma)$ by an effective Lorentz
factor $\gamma_{eff}(t) = \gamma (\vec{v}(t)) \neq const$, which varies with the
changes of velocity $\vec{v}(t)$.
Similar analysis shows
that the assumption $\vec{p} = const$ leads to the conclusion analogous to
that resulting from the assumption $E_{\phi} = const$
that the velocity $\vec{v}$
of the moving quantum unstable systems can not be constant in time.
This is the consequence
of the relativistic formula for the momentum $\vec{p}$.

From results presented in Fig. (\ref{f2}) it is seen that with increasing time, $t$,
the amplitude of fluctuations of $m_{\phi}(t)$ grows. So according to (\ref{E=mg(v)}), in order to compensate
these growing fluctuations,  the fluctuations of the velocity $\vec{v}(t)$ of the unstable system have to grow.
This means that with increasing time, $t$, (at $t > \tau_{\phi}$), deviations of
the decay law of moving unstable system from the classical relation ${\cal P}_{0}(t/\gamma)$
should be more visible and should grow. This effect explains the results presented in \cite{ku-2014}
where with the increasing time
the increasing difference between ${\cal P}_{0}(t/\gamma)$ and ${\cal P}_{p}(t)$ was indicated and analyzed.

One more remark.
Let us denote by ${\cal O}'$
the reference frame
which moves together with the moving quantum unstable system considered and
in which this system is in rest.
This reference frame
moves relative to ${\cal O}$ with the velocity $\vec{v}$.
The property that the velocity $\vec{v}$ of
the moving quantum unstable system
can not be
constant in time has an effect that $\frac{d \vec{v}}{dt} \neq 0$.
Therefore the rest reference frame ${\cal O}'$ of
such a system
can not be the inertial one.
This observation means that there does not exist a Lorentz transformation describing a transition
from the inertial rest reference frame ${\cal O}$ of the observer into the the noninertial
rest reference frame ${\cal O}'$ of the moving quantum unstable system.

The last remark.
It seems that the above-described effect can be relatively easily verified experimentally.
It is because the conclusion that the velocity $\vec{v}$ of the moving quantum unstable system must vary in time means
that $\frac{d \vec{v}}{dt} \neq 0$.
Therefore
the moving freely charged unstable particles
(or neutral unstable particles with nonzero magnetic moment) should emit
electromagnetic radiation of the very broad spectrum: From very small  up to extremely large frequencies
(see \cite{ku1-2014}).
Thus this effect can be verified by using
currently carried out experiments, which use a
beam of charged unstable particles (eg. $\pi^{\pm}$ mesons or muons) or ions of radioactive elements moving along a straight line.
A section of the track of these particles, where they are moving freely, should be surrounded by sensitive antennae connected
to the receivers being
able to register that a broad spectrum of the electromagnetic radiation.  Then every signal coming from the beam registered
by these receivers will be the proof that the above described effect takes place.\\

\noindent
{\bf Acknowledgments:}
The authors would like to thank E. V. Stefanovich
for valuable comments
The work was supported by the Polish  NCN grant No
DEC-2013/09/B/ST2/03455.

\end{document}